\documentstyle[12pt]{article}
\newlength{\extraspace}
\newlength{\extraspaces}
\newcommand{\be}{\begin{equation}
\addtolength{\abovedisplayskip}{\extraspaces}
\addtolength{\belowdisplayskip}{\extraspaces}
\addtolength{\abovedisplayshortskip}{\extraspace}
\addtolength{\belowdisplayshortskip}{\extraspace}}
\newcommand{\ee}{\end{equation}}

\newcommand{\ba}{\begin{eqnarray}
\addtolength{\abovedisplayskip}{\extraspaces}
\addtolength{\belowdisplayskip}{\extraspaces}
\addtolength{\abovedisplayshortskip}{\extraspace}
\addtolength{\belowdisplayshortskip}{\extraspace}}
\newcommand{\ea}{\end{eqnarray}}

\newcommand{\nonu}{\nonumber \\[.5mm]}
\newcommand{\A}{&\!\!\!}

\newcommand{\newsection}[1]{
\vspace{7mm} \pagebreak[3] \addtocounter{section}{1}
\setcounter{subsection}{0} \setcounter{footnote}{0}
\begin{center}
{\large {\bf \thesection. #1}}
\end{center}
\nopagebreak
\medskip
\nopagebreak \hspace{3mm}}

\begin{document}
\begin{center}
{{\bf On the Regularization  of Kerr-NUT spacetime: I}}
\end{center}
\centerline{Gamal G.L. Nashed}
\bigskip

\centerline{\it Mathematics Department, Faculty of Science, King
Faisal University, P.O. Box.} \centerline{\it 380 Al-Ahsaa 31982,
the Kingdom of Saudi Arabia \footnote{ Mathematics Department,
Faculty of Science, Ain
Shams University, Cairo, 11566, Egypt. \vspace*{0.4cm}\\
Center for Theoretical Physics, British University of  Egypt
 Sherouk City 11837, P.O. Box 43, Egypt.
\vspace*{0.4cm} \\
 Egyptian Relativity Group (ERG) URL:
http://www.erg.eg.net.}}

\bigskip
\centerline{ e-mail: nashed@bue.edu.eg}

\hspace{2cm} \hspace{2cm}
\\
\\
\\
\\
\\
\\
\\
\\
\\

Within the framework of teleparallel equivalent of  general
relativity (TEGR) theory, calculation of the total energy and
momentum of Kerr-NUT spacetimes have been employed using two
methods of the gravitational energy-momentum, which is coordinate
independent, and the Riemannian connection  1-form,
${\widetilde{\Gamma}_\alpha}^\beta$. It has been shown that the
two methods give the same an unacceptable result, i.e., divergent
value. Therefore, a local Lorentz transformation that plays a role
of a regularizing tool, which subtracts the inertial effects
without distorting the true gravitational contribution, has been
suggested. This transformation  keeps the resulting spacetime to
be a solution of the equations of motion of TEGR.

\newsection{Introduction}

 The geometry of Einstein general relativity (GR) is based on  the
Riemannian geometry  which  can define  both metric and connection
uniquely. This geometry is the main reason responsible for the
problem of defining a consistent expression of energy in GR in
addition to the equivalence principle of the gravitational theory.
Using the Lagrange-Noether approach, one can derive  the conserved
currents that  arise from the invariance of the classical action
under transformations of fields. In Riemannian geometry one can
not find symmetries that can be used to generate the conserved
energy-momentum currents, but one can only speak about the energy
of asymptotically  flat spacetime. Earlier analysis of this
problem can be found in  details  in [ references 1)$\sim5)$ and
references therein].

As is well known, there are two ways to describe the gravitational
interaction: one by curvature and the other by torsion.$^{ \rm
{6)}}$  According to GR, curvature is due to geometry
 from which a successful description of the gravitational
interaction is carried out. On the other hand, teleparallelism
 attributes gravitation to torsion. In this case, torsion accounts
for gravitation not by geometrizing the interaction, but by acting
as a force. Accordingly, in the teleparallel equivalent of GR,
there are no geodesics, but force equations are quite analogous to
the Lorentz force equation of electrodynamics.$^{ \rm {7)}}$
Therefore, gravitational interaction, can be described either in
terms of curvature, as is usually conducted in GR, or in terms of
torsion, in which case we have the teleparallel gravity.

Teleparallel theories are interesting for many reasons: First  GR
can be viewed as a particular theory of teleparallelism and, thus,
teleparallelism could be considered at the very least as a
different point of view that can lead to the same results.$^{ \rm
{8)}}$  Second, within this framework, one can define an
energy-momentum tensor for the gravitational field that is a true
tensor under all general coordinate transformations. This is the
reason why teleparallelism was reconsidered by M\o ller  when he
was studying the problem of defining an energy-momentum tensor for
the gravitational field.$^{ \rm {9)}}$  The idea was taken over by
Pellegrini and Pleba\'{n}ski who constructed the general
Lagrangian for these theories.$^{ \rm {10)}}$  The third reason as
to  why these theories are interesting is that they can be seen as
gauge theories of the translation group (not the full
Poincar$\acute{e}$ group) and, thus, they give an alternative
interpretation of GR.$^{ \rm {11),\ 12)}}$  Teleparallel theory
 is considered as an essential part of
generalized \ non-Riemannian \ theories such as the \
Poincar$\acute{e}$ gauge \ theory$^{ \rm {13)\sim 20)}}$ or
metric-affine gravity.$^{ \rm {11),\ 21)\sim 25)}}$   Within the
framework of metric-affine gravity, a stationary axially symmetric
exact solution of the vacuum field equation is obtained.$^{ \rm
{26)}}$

Geometrically, teleparallel models are described by the
Weitzenb$\ddot{o}$ck spacetime  which is characterized by the
vanishing curvature (constructed from the connection of this
geometry) and non-zero torsion. The tetrad (or a coframe)
 is the basic field variable which can be
treated as the gauge potential corresponding to the group of local
translations. Then the torsion is naturally interpreted as the
corresponding gauge field strength. As a result, a gravitational
teleparallel Lagrangian is straightforwardly constructed, in a
Yang-Mills manner, from the quadratic torsion invariants.

Mathematically, there are infinitely many tetrads since a
reference frame of an observer can obviously be constructed
infinitely. In particular, from a given tetrad field ${h_i}^\mu$,
we can obtain a continuous family of tetrads by performing the
local Lorentz transformation ${h'_i}^\alpha={\Lambda^\alpha}_\beta
{h_i}^\beta $ , where the elements of the Lorentz matrix
${\Lambda^\alpha}_\beta(x)$ are arbitrary functions of the
spacetime coordinates.$^{ \rm {27)}}$

An important difference between Einstein GR theory and
teleparallel theories  is that it is possible to distinguish
between gravitation and inertia.$^{ \rm {27),\ 28)}}$  Since
inertia is in the realm of the pseudotensor behavior of the usual
expressions for the gravitational energy-momentum density, it
seems possible in teleparallel gravity to write down a tensorial
expression for such density.$^{ \rm {29)}}$  With the purpose of
getting a deeper insight into the covariant teleparallel
formalism, as well as to test how it works Lucas  et al.$^{ \rm {
27)}}$ reanalyze the computation of the total energy of two
examples. Recently Obukhov et al.$^{ \rm { 30)}}$ computed the
energy and momentum transported by exact plane gravitational-wave
solutions of Einstein equations using the teleparallel equivalent
formulation of Einstein's theory. {\it It is our aim in this
current work to calculate energy and momentum of Kerr-NUT
spacetime using two methods: the gravitational energy-momentum
which is a coordinate independent and  the Riemannian connection
1-form, ${\widetilde{\Gamma}_\alpha}^\beta$. The value of energy
of Kerr-NUT spacetime has been shown to have a divergent result.
Therefore,  an appropriate Local Lorentz transformation has been
introduced. Using this transformation,  it has been shown that the
value of energy is always acceptable}.

In \S 2, a brief review of the derivation of the TEGR field
equations  is given. A summary of the derivation of energy and
angular momentum using the Hamiltonian formulation in TEGR is also
given in \S 2.  In \S 3,  Kerr-NUT spacetime and calculation of
its energy,  using the definition given in \S 2 has been presented
 and a divergent value is obtained. To make the picture  clearer
we use another definition to calculate the energy. Therefore, we
give a brief review of the covariant formalism for the
gravitational energy-momentum  which is described by the pair $(
\vartheta^{\alpha}, {\Gamma_\alpha}^\beta)$ in \S 3.  Using the
Riemannian connection  1-form, ${\widetilde{\Gamma}_\alpha}^\beta$
we repeat the calculation of energy and obtain the same divergent
value. In \S 4,  explicate and adequate
 calculations show that due to an inconvenient choice of a reference
system, traditional computation of the total energy of Kerr-NUT
spacetime are always divergent! Therefore,  a new local Lorentz
transformation which when applied to the Kerr-NUT spacetime and
repeat the calculation either using the gravitational
energy-momentum or  the Riemannian connection 1-form
${\widetilde{\Gamma}_\alpha}^\beta$, we get  finite and acceptable
result. In section \S 5, we give another Kerr-NUT spacetime and
calculate the energy using the two definitions and got a non
acceptable result. Using the local Lorentz transformation
suggested in \S 4, we get an acceptable result obtained in \S 4.
Final section is
devoted to main result and discussion.\vspace{0.4cm}\\
\newsection{The  TEGR for gravitation}
In a spacetime with absolute parallelism the parallel vector field
${h_a}^\mu$ defines the nonsymmetric affine connection \be
{\Gamma^\lambda}_{\mu \nu} \stackrel{\rm def.}{=} {h_a}^\lambda
{h^a}_{\mu, \nu}, \ee where $h_{a \mu, \nu}=\partial_\nu h_{a
\mu}$\footnote{spacetime indices $\mu, \ \ \nu, \cdots$ and
SO(3,1) indices a, b $\cdots$ run from 0 to 3. Time and space
indices are indicated to $\mu=0, i$, and $a=(0), (i)$.}. The
curvature tensor defined by ${\Gamma^\lambda}_{\mu \nu}$ is
identically vanishing, however. The metric tensor $g_{\mu \nu}$
 is given by
 \be g_{\mu \nu}= O_{a b} {h^a}_\mu {h^b}_\nu, \ee with the
Minkowski metric $O_{ a b}=\textrm {diag}(+1\; ,-1\; ,-1\; ,-1)$
\footnote{ Latin indices are rasing and lowering with the aid of
$O_{ a b}$ and $O^{ a b}$.}.

  The Lagrangian density for the gravitational field in the TEGR,
  in the presence of matter fields, is given by$^{ \rm {31)}}$\footnote{Throughout this paper we use the relativistic
units$\;$ , $c=G=1$ and $\kappa={8\pi}$.}
 \be  {\cal L}_G  =  \sqrt{-g} L_G =- \displaystyle {\sqrt{-g}
\over 16\pi}  \left( \displaystyle {T^{abc}T_{abc} \over
4}+\displaystyle {T^{abc}T_{bac} \over 2}-T^aT_a
  \right)-L_m= - \displaystyle {\sqrt{-g} \over 16\pi} {\Sigma}^{abc}T_{abc}-L_m,\ee
where $g=det(g_{\mu \nu})$. The tensor ${\Sigma}^{abc}$ is defined
by \be {\Sigma}^{abc} \stackrel {\rm def.}{=} \displaystyle{1
\over 4}\left(T^{abc}+T^{bac}-T^{cab}\right)+\displaystyle{1 \over
2}\left(O^{ac}T^b-O^{ab}T^c\right).\ee $T^{abc}$ and $T^a$ are the
torsion tensor and the basic vector field  defined by \be
{T^a}_{\mu \nu} \stackrel {\rm def.}{=}
{h^a}_\lambda{T^\lambda}_{\mu
\nu}=\partial_\mu{h^a}_\nu-\partial_\nu{h^a}_\mu,\ee and \be
 T^\mu \stackrel {\rm def.}{=} {{T^\nu}_\nu}^\mu, \qquad T^a \stackrel {\rm
def.}{=} {h^a}_\mu T^\mu={{T^b}_b}^a.\ee  The quadratic
combination $\Sigma^{abc}T_{abc}$ is proportional to the scalar
curvature $R$, except for a total divergence term.$^{ \rm {32)}}$
$L_m$ represents the Lagrangian density for matter fields.

The gravitational field equations for the system described by
${\it L_G}$ are the following
 \be h_{a \lambda}h_{b \mu}\partial_\nu\left(\sqrt{-g}{\Sigma}^{b \lambda \nu}\right)-\sqrt{-g}\left(
 {{\Sigma}^{b \nu}}_a T_{b \nu \mu}-\displaystyle{1 \over 4}h_{a \mu}
 T_{bcd}{\Sigma}^{bcd}\right)= \displaystyle{1 \over 2}{\kappa} \sqrt{-g}T_{a
 \mu},\ee

where \[ \displaystyle{ \delta L_m \over \delta h^{a \mu}} \equiv
\sqrt{-g} T_{a \mu}.\] It  is possible to prove by explicit
calculations that the left hand side of the symmetric part of the
field equations (7) is exactly given by$^{ \rm { 31)}}$
 \[\displaystyle{\sqrt{-g} \over 2} \left[R_{a
\mu}-\displaystyle{1 \over 2}h_{a \mu}R \right]. \] The
axial-vector part of the torsion tensor $A_\mu$ is defined by \be
A_\mu \stackrel{\rm def.}{=} {1 \over 6} \epsilon_{\mu \nu \rho
\sigma} T^{\nu \rho \sigma}={1 \over 3} \epsilon_{\mu \nu \rho
\sigma} \gamma^{\nu \rho \sigma}, \qquad where \qquad
\epsilon_{\mu \nu \rho \sigma} \stackrel{\rm def.}{=} \sqrt{-g}
\delta_{\mu \nu \rho \sigma}, \ee with $\gamma_{\nu \rho
\sigma}=O^{a b}h_{a \nu} h_{b \rho\; ;\; \sigma}$ being the
contortion tensor and $\delta_{\mu \nu \rho \sigma}$ is completely
antisymmetric and normalized as $\delta_{0123}=-1$.

The definition of the gravitational energy-momentum $P^a$
four-vector  has the form$^{ \rm {32)}}$ \be P^a\stackrel{\rm
def.}{=}-\int_V d^3 x
\partial_i \Pi^{ai},\ee where $V$ is an arbitrary volume of the
three-dimensional space. In the configuration space we have \ba \A
\A \Pi^{ai} \stackrel{\rm def.}{=} -\frac{2}{\kappa} \sqrt{-g}
\Sigma^{a0i}, \quad with \quad
\partial_\nu(\sqrt{-g}\Sigma^{a \lambda \nu})\stackrel{\rm
def.}{=}\displaystyle{\kappa \over 2}\sqrt{-g}{h^a}_\mu
(t^{\lambda \mu}+T^{\lambda \mu}), \nonu
\A \A  where \quad   t^{\lambda \mu}\stackrel{\rm
def.}{=}\frac{1}{2\kappa} \left(4\Sigma^{bc
\lambda}{T_{bc}}^\mu-g^{\lambda \mu} \Sigma^{bcd}T_{bcd}
\right).\ea

 Maluf et al.$^{ \rm {31), \ 32)}}$
 defined \be L^{ab} = 2\int_V d^3x M^{[ab]}, \ee  as the
4-angular-momentum of the gravitational field for an arbitrary
volume V of the three-dimensional space.
\newsection{First Kerr-NUT spacetime}
The covariant form of the first Kerr-NUT tetrad field having
axial symmetry in spherical coordinates $(t,r,\theta, \phi)$, can
be written as
 \be
\left( {h^\alpha}_i \right)_1= \left( \matrix{ {\cal A}_1 & 0 & 0
& 0\vspace{3mm} \cr 0 & {\cal A}_2 & 0 & 0 \vspace{3mm} \cr 0 & 0
& {\cal A}_3 & 0 \vspace{3mm}  \cr {\cal A}_4
 & 0 &0& {\cal A}_5\sin\theta \cr } \right), \ee
 where  ${\cal A}_i$, $i=1\cdots 5$ are  functions of
 $r$ and $\theta$  having the form  \ba
{\cal A}_1 \A=\A -\sqrt{\frac{{\cal A}{\cal B}}{\cal
C}}\sin\theta, \qquad {\cal A}_2=\sqrt{\frac{\cal A}{\cal B}},
\qquad
  {\cal A}_3=\sqrt{\cal A},  \qquad {\cal A}_4=\frac{{\cal G}}{\sqrt{\cal {A  C}}},   \qquad {\cal A}_5=\sqrt{\frac{\cal C}{\cal A}},
 \nonu
\qquad {\cal A}\A=\A{r^2+\left(L+a\cos\theta\right)^2}, \qquad
{\cal B } = {r^2-2Mr+a^2-L^2}, \nonu
 {\cal C}\A=\A a{\cal B }\cos^3\theta(a\cos\theta-L)-\cos^2\theta\Biggl(r^4+8a^2L^2+6r^2L^2-a^4-3L^4-8MrL^2+4Mra^2\Biggr)\nonu
 \A \A +4aL{\cal B }\cos\theta +2Mra^2+L^4+r^4+r^2a^2+2r^2L^2+3a^2L^2,\nonu
 {\cal G} \A=\A -2({\cal B } L\cos\theta +a[L^2+Mr]\sin^2\theta), \ea  where  $M$ , $a$  and $L$ are the mass,
  the rotation  and the NUT parameters respectively.$^{ \rm {33)}}$  We consider a non asymptotically flat spacetime
in this paper, and impose the boundary condition that for $r
\rightarrow \infty$ and $L\rightarrow 0$
 the tetrad (12) approaches the tetrad of Minkowski spacetime,  in Cartesian coordinate.
 The metric tensor $g_{i j}\stackrel {\rm def.}{=}  O_{\mu \nu} {h^\mu}_i {h^\nu}_j$ associated with the tetrad field (12) has the form
 \be
ds^2= \left({{\cal A}_1}^2-{{\cal A}_4}^2\right) dt^2 -{{\cal
A}_2}^2  dr^2 -{{\cal A}_3}^2 d\theta^2-{{\cal A}_5}^2\sin^2\theta
d\phi^2- {\cal A}_5{\cal A}_4\sin\theta dt d\phi, \ee which is the
Kerr-NUT  spacetime written in Boyer-Lindquist coordinates.$^{ \rm
{27)}}$

The previously obtained solutions, Schwarzschild and Kerr
spacetimes can be generated as special solutions of the tetrad
(12) using (13) by putting $a=0$, $L=0$ and  $L=0$
respectively.$^{ \rm {34), \ 35)}}$

Now we are going to calculate the energy content of the tetrad
field  (12) using  (13). The non-vanishing components of the
tensor $\Sigma^{a b c}$ needed to the calculation of energy have
the form \ba \Sigma^{3 0 1}\A \cong
\A\frac{L\cos\theta}{r^3\sin^2\theta}+O\left(\frac{1}{r^4}\right),\qquad
\Sigma^{4 0 1}
 \cong \frac{1}{r^3\sin^2\theta }\Biggl\{2r^2\sin^2\theta-2L^2(1+\cos^2\theta)\nonu
\A \A
-4La\cos\theta\sin^2\theta+a^2\sin^2\theta(1-3\cos^2\theta)\Biggr\}+O\left(\frac{1}{r^4}\right).
\ea Using Eq. (15) in Eq. (9) we finally obtain \ba \A \A
P^{(0)}=E=-\oint_{S \rightarrow \infty}
 dS_k \Pi^{(0) k}=-\displaystyle {1  \over 4 \pi} \oint_{S \rightarrow \infty}
 dS_k  \sqrt{-g}
{h^{(0)}}_\mu {\Sigma}^{\mu 0 k}  \cong \infty!\ea It follows from
Eq. (16) that the energy of the Kerr-Tube-NUT spacetime is
divergent which is not an  acceptable result.

Due to the above non physical result and to make the picture more
clear we will use another method
 to calculate the energy of tetrad (12) to show if the divergent result will continue or not?

\centerline{\bf Notation}

We use the Latin indices ${\it i, j, \cdots }$ for local holonomic
spacetime coordinates and the Greek indices $\alpha$, $\beta$,
$\cdots$ label (co)frame components. Particular frame components
are denoted by hats, $\hat{0}$,$\hat{1}$, etc. As usual, the
exterior product is denoted by $\wedge$, while the interior
product of a vector $\xi$ and a p-form $\Psi$ is denoted by $\xi
\rfloor \Psi$. The vector basis dual to the frame 1-forms
$\vartheta^{\alpha}$ is denoted by $e_\alpha$ and they satisfy
$e_\alpha \rfloor \vartheta^{\beta}={\delta}_\alpha^\beta$. Using
local coordinates $x^i$, we have $\vartheta^{\alpha}=h^\alpha_i
dx^i$ and $e_\alpha=h^i_\alpha \partial_i$ where $h^\alpha_i$ and
$h^i_\alpha $ are the covariant and contravariant components of
the tetrad field. We define the volume 4-form by $\eta \stackrel
{\rm def.}{=} \vartheta^{\hat{0}}\wedge \vartheta^{\hat{1}}\wedge
\vartheta^{\hat{2}}\wedge\vartheta^{\hat{3}}.$  Furthermore, with
the help of the interior product we define \[\eta_\alpha \stackrel
{\rm def.}{=} e_\alpha \rfloor \eta = \ \frac{1}{3!} \
\epsilon_{\alpha \beta \gamma \delta} \ \vartheta^\beta \wedge
\vartheta^\gamma \wedge \vartheta^\delta,\]  where
$\epsilon_{\alpha \beta \gamma \delta}$ is completely
antisymmetric tensor with $\epsilon_{0123}=1$. \[\eta_{\alpha
\beta} \stackrel {\rm def.}{=} e_\beta \rfloor \eta_\alpha =
\frac{1}{2!}\epsilon_{\alpha \beta \gamma \delta} \
\vartheta^\gamma \wedge \vartheta^\delta,\qquad \qquad
\eta_{\alpha \beta \gamma} \stackrel {\rm def.}{=} e_\gamma
\rfloor \eta_{\alpha \beta}= \frac{1}{1!} \epsilon_{\alpha \beta
\gamma \delta} \ \vartheta^\delta,\]  which are bases for 3-, 2-
and 1-forms respectively. Finally, \[\eta_{\alpha \beta \mu \nu}
\stackrel {\rm def.}{=} e_\nu \rfloor \eta_{\alpha \beta \mu}=
e_\nu \rfloor e_\mu \rfloor e_\beta \rfloor e_\alpha \rfloor
\eta,\] is the Levi-Civita tensor density. The $\eta$-forms
satisfy the useful identities: \ba \vartheta^\beta \wedge
\eta_\alpha \A \stackrel {\rm def.}{=}  \A \delta^\beta_\alpha
\eta, \qquad \vartheta^\beta \wedge \eta_{\mu \nu}  \stackrel {\rm
def.}{=} \delta^\beta_\nu \eta_\mu-\delta^\beta_\mu \eta_\nu,
\qquad  \vartheta^\beta \wedge \eta_{\alpha \mu \nu}  \stackrel
{\rm def.}{=}  \delta^\beta_\alpha \eta_{\mu \nu}+\delta^\beta_\mu
\eta_{\nu \alpha}+\delta^\beta_\nu \eta_{ \alpha \mu}, \nonu
\vartheta^\beta \wedge \eta_{\alpha \gamma \mu \nu}  \A \stackrel
{\rm def.}{=}  \A \delta^\beta_\nu \eta_{\alpha \gamma
\mu}-\delta^\beta_\mu \eta_{\alpha \gamma \nu
}+\delta^\beta_\gamma \eta_{ \alpha \mu \nu}-\delta^\beta_\alpha
\eta_{ \gamma \mu \nu}. \ea The line element $ds^2 \stackrel {\rm
def.}{=} g_{\alpha \beta} \vartheta^\alpha \bigotimes
\vartheta^\beta$ is defined by the spacetime metric $g_{\alpha
\beta}$.

Teleparallel geometry can be viewed as a gauge theory of
translation.$^{ \rm {11), \ 12), \ 36)\sim 41)}}$ In this geometry
the coframe $\vartheta^\alpha$  plays the role of the gauge
translational potential of the gravitational field.  GR can be
reformulated as the teleparallel theory. Geometrically,
teleparallel gravity can be considered as a special case
 of the metric-affine gravity in which the coframe 1-form
  $\vartheta^\alpha$ and the local Lorentz connection  are subject to the distant parallelism
constraint ${R_\alpha}^\beta=0$.$^{ \rm {42), \ 43), \ 44)}}$  In
this geometry the torsion 2-form \be
T^\alpha=D\vartheta^\alpha=d\vartheta^\alpha+{\Gamma_\beta}^\alpha\wedge
\vartheta^\beta=\frac{1}{2}{T_{\mu \nu}}^\alpha \vartheta^\mu
\wedge \vartheta^\nu=\frac{1}{2}{T_{i j}}^\alpha dx^i \wedge
dx^j,\ee arises as the gravitational gauge field strength,
${\Gamma_\alpha}^\beta$ being the Weitzenb$\ddot{o}$ck connection
1-form, $d$ is the exterior derivative and $D$ is the exterior
covariant derivative. The torsion $T^\alpha$ can be decomposed
into three irreducible pieces: the tensor part, the trace, and the
axial trace, given respectively by$^{ \rm {12), \ 27)}}$, for
example \ba {^ {\tiny{( 1)}}T^\alpha}  \A \stackrel {\rm def.}{=}
\A T^\alpha-{^ {\tiny{( 2)}}T^\alpha}-{^ {\tiny{( 3)}}T^\alpha},
\qquad with \nonu
{^  {\tiny{( 2)}}T^\alpha}  \A \stackrel {\rm def.}{=} \A
\frac{1}{3} \vartheta^\alpha\wedge T, \quad where \quad T=
\left(e_\beta \rfloor T^\beta\right), \qquad e_\alpha \rfloor
T={T_{\mu \alpha}}^\mu, \quad vectors \ of \  trace \ of \ torsion
\nonu
{^  {\tiny{( 3)}}T^\alpha}  \A \stackrel {\rm def.}{=} \A
\frac{1}{3} e^\alpha\rfloor P, \quad with \quad
P=\left(\vartheta^\beta \wedge T_\beta\right), \quad
e_\alpha\rfloor P=T^{\mu \nu \lambda}\eta_{\mu \nu \lambda
\alpha}, \quad axial \ of \  trace \ of \ torsion.\nonu
\A \A \ea The Lagrangian of the teleparallel equivalent using the
language of forms has the form, \{The effect of adding the
non-Riemannian parity odd pseudoscalar curvature to the
Hilbert-Einstein-Cartan scalar curvature was studied by many
authors cf., Ref. 45) and references therein.\},  \be V=
-\frac{1}{2\kappa}T^\alpha \wedge ^\ast \left({^ {\tiny{(
1)}}T_\alpha}-2{^  {\tiny{( 2)}}T_\alpha} -\frac{1}{2}{^ {\tiny{(
3)}}T_\alpha} \right). \ee $\kappa=8\pi G/c^3$, $G$ is the Newton
gravitational constant, $c$ is the speed of light and $\ast$
denotes the Hodge duality in the metric $g_{\alpha \beta}$ which
is assumed to be flat Minkowski metric $g_{\alpha \beta}=O_{\alpha
\beta}=diag(+1,-1,-1,-1)$, that is used to raise and lower local
frame (Greek) indices.

The variation of the total action with respect to the coframe
gives the field equations in the  from$^{ \rm {27)}}$ \be
DH_\alpha-E_\alpha=\Sigma_\alpha, \ \ where \ \ \Sigma_\alpha
\stackrel {\rm def.}{=} \frac{\delta L_{mattter}}{\delta
\vartheta^\alpha},\ee  is the canonical energy-momentum current
3-form of matter  which is considered as the source. In accordance
with the general Lagrange-Noether scheme$^{ \rm {11), \ 38)}}$ one
derives from (20) the translational momentum 2-form and the
canonical energy-momentum 3-form: \be H_{\alpha} \stackrel {\rm
def.}{=} -\frac{\partial V}{\partial T^\alpha}=\frac{1}{\kappa}
\ast \left({^  {\tiny{( 1)}}T_\alpha}-2{^  {\tiny{( 2)}}T_\alpha}
-\frac{1}{2}{^  {\tiny{( 3)}}T_\alpha} \right), \qquad E_\alpha
\stackrel {\rm def.}{=} \frac{\partial V}{\partial
\vartheta^\alpha}=e_\alpha \rfloor V+\left(e_\alpha \rfloor
T^\beta \right) \wedge H_\beta. \ee Due to geometric identities$^{
\rm {43)}}$, the Lagrangian (20) can be recast as \be
V=-\frac{1}{2}T^\alpha\wedge H_\alpha.\ee The presence of the
connection field ${\Gamma^\alpha}_\beta$ plays an important
regularizing
role due to the following: \vspace{.3cm}\\
\underline{i}: The theory becomes explicitly covariant under the
local Lorentz transformations of the coframe, i.e.,
 the Lagrangian (20) is invariant under the change of
variables \be   \vartheta'^\alpha={\Lambda^\alpha}_\beta
\vartheta^\beta, \qquad
{\Gamma'_\alpha}^\beta={\Lambda^\mu}_\alpha {\Gamma_\mu}^\nu
{(\Lambda^{-1})^\beta}_\nu-{(\Lambda^{-1})^\beta}_\gamma
d{\Lambda^\gamma}_\alpha.\ee Due to the non-covariant
transformation law  of  ${\Gamma_\alpha}^\beta$
 as shown by Eq. (24), if a connection
vanishes in a given frame, it will not vanish in any other frame
related to the first by a
local Lorentz transformation.\vspace{0.5cm}\\
 \underline{ii}: ${\Gamma_\alpha}^\beta$ plays an
important role in the teleparallel framework. This role represents
the inertial effects which arise from the choice of the reference
system$^{ \rm {27)}}$. The  contributions of this inertial in many
cases lead to unphysical results for the total energy of the
system. Therefore, the role of the teleparallel connection is to
separate the inertial contribution from the truly gravitational
one. Since the teleparallel curvature is zero, the connection is a
"pure gauge", that is \be
{\Gamma_\alpha}^\beta={(\Lambda^{-1})^\beta}_\gamma d
{\Lambda^\gamma}_\alpha.\ee The  Weitzenb$\ddot{o}$ck connection
always has the form (25). The   translational momentum has the
form$^{ \rm {27)}}$ \be
\widetilde{H}_\alpha=\frac{1}{2\kappa}{\widetilde{\Gamma}}^{\beta
\gamma}\wedge  \eta_{\alpha \beta \gamma}, \qquad
{\Gamma_\alpha}^\beta \stackrel {\rm def.}{=} {\widetilde
{\Gamma}_\alpha}^\beta -{K_\alpha}^\beta,\ee  with ${\widetilde
{\Gamma}_\alpha}^\beta $  is the purely Riemannian connection and
$K^{\mu \nu}$ is the contorsion 1-form which is related to the
torsion through the relation

\be T^\alpha  \stackrel {\rm def.}{=}  {K^\alpha}_\beta \wedge
\vartheta^\beta.\ee

 Using the spherical local coordinates
$(t,r,\theta, \phi)$ the Kerr-NUT, using Eq. (12), frame is
described by the coframe components:
 \ba
{\vartheta_1}^{\hat{0}}\A=\A {\cal A}_1
 cdt, \qquad
{\vartheta_1}^{\hat{1}}={\cal A}_2 dr, \qquad
{\vartheta_1}^{\hat{2}}={\cal A}_3 d\theta, \qquad
{\vartheta_1}^{\hat{3}}={\cal A}_4dt+{\cal A}_5 \sin\theta d\phi.
\ea
   If we take coframe  (28), as
well as the Riemannian connection ${\tilde{\Gamma}_\alpha}^\beta$
and substitute into (26) we finally get\footnote{$\cdots$ means
terms which are multiply by $d\theta\wedge dr$, $d\theta\wedge
dt$, $dr\wedge d\phi$ ect.} \ba \widetilde{H}_{\hat{0}} \A \cong
\A \frac{- \sin\theta }{8r^2\pi}
\Biggl[2Mr^2-2r^3-2Ma^2\cos^2\theta-rM^2+M^3-ra^2\sin^2\theta+2rL^2+(2raL-6LMa)\cos\theta\nonu
\A \A +4ML^2\cot^2\theta\Biggr]d\theta\wedge d\phi
 +\cdots+O\left(\frac{1}{r^2}\right).\ea
Using Eq. (29) to compute the total energy at a fixed time in the
3-space with a spatial  2-dimensional boundary surface $\partial S
=\{r = R, \theta,\phi\}$ we finally obtain \be \widetilde{E}
=\int_{\partial S} \widetilde{H}_{\hat{0}}=\infty!\ee which is
identical with Eq. (16).
\newsection{On the choice of the frame}
Let us consider the Lorentz transformation described by the matrix
 \be \left({\Lambda^\alpha}_{  \beta} \right)
= \left( \matrix{ B_1 &  B_2 & B_3 &  B_4
 \vspace{3mm} \cr  C_1 \sin\theta
\cos\phi  & C_2 \sin\theta \cos\phi & C_3 \cos\theta \cos\phi
 & C_4 \sin\phi  \sin\theta \vspace{3mm} \cr
  F_1 \sin\theta \sin\phi  & F_2  \sin\theta \sin\phi &F_3
   \cos\theta
\sin\phi & F_4 \cos\phi \sin\theta \vspace{3mm} \cr
 G_1\cos\theta&   G_2 \cos\theta & G_3\sin\theta  &
   G_4 \cos\theta  \cr }
\right)\; , \ee where $B_i\; ,  \ C_i\; , \ F_i \ \ and \ \  G_i\;
, \
 i=1 \cdots 4 $ are defined as:
 \ba B_1 \A
=\A \frac{\sin\theta}{\sqrt{{\cal AB}L_5}}\Biggl(a
L^3\cos\theta-a^2L^2(2-\cos^2\theta)+L^2(Mr-r^2)-aL(3r^2-4Mr-3a^2)\cos\theta-a^4\cos^2\theta\nonu
\A \A
-r^2a^2(1+\cos^2\theta)-r^4-a^2rM+r^3M+2Ma^2r\cos^2\theta\Biggr)\;
, \qquad B_2=-\displaystyle{(M r+LL_1)    \over \sqrt{\cal AB}}\;
,
 \quad B_3=0\; , \nonu
 B_4 \A =\A \displaystyle{ M r \chi+L(aL_1[1+ \cos^2\theta]+2r^2\cos\theta)  \over \sqrt{{\cal A}L_5}}\; , \quad C_1 = \displaystyle{( -2r_1(L{\cal B}\cos\theta+ M r \chi+aL^2\sin^2\theta)\sin\phi+L_3\cos\phi) \over \sqrt{{\cal AB}L_5}}\; ,\nonu
  C_2 \A =\A \displaystyle {-1 \over  \sqrt{\cal AB} \cos \phi} \left(r_1\alpha
- [ M r+LL_1] \cos \phi \right)\; ,\quad
  C_3= \displaystyle{\alpha \over {\cal A}\cos \phi} \; , \qquad
  C_4 =\frac{1}{\sqrt{{\cal A}L_5} \sin \phi}\left( \displaystyle{ {  r_1{\cal A}\sin\phi-L_4\cos\phi
  }}\right)\; , \nonu
F_1 \A =\A\displaystyle{( 2r_1(L{\cal B}\cos\theta- M r
\chi-aL^2\sin^2\theta)\cos\phi+L_3\sin\phi) \over \sqrt{{\cal
AB}L_5}}\; , \quad F_2=\displaystyle\displaystyle {-1 \over
\sqrt{\cal AB} \sin \phi}\left(r_1\beta- [ M r+LL_1] \sin \phi
\right)\; ,\nonu
 F_3\A =\A \displaystyle{\beta \over {\cal A}\sin \phi}\; , \quad
 F_4=\frac{1}{\sqrt{{\cal A}L_5} \sin \phi}\left( \displaystyle{ {-  r_1{\cal A}\cos\phi+L_4\sin\phi
  }}\right)\; , \quad
  G_1 = \displaystyle{-( M r+LL_1)(r^2+a^2+L^2) \over  \sqrt{{\cal AB}L_5} }\; ,\nonu
G_2 \A =\A \displaystyle{( M r-r^2-a^2-aL\cos\theta) \over
\sqrt{\cal AB} }\;, \qquad  G_3 = -\frac{r_1}{{\cal A}}\; ,\nonu
 G_4 \A =\A  \displaystyle {( -M r\chi +L[2L^2\cos\theta-aL(1-3\cos^2\theta)-a^2\cos\theta\sin^2\theta])
  \over  \sqrt{{\cal A}L_5}}\; ,
 \ea
 where $\Omega, \ \  \Upsilon, \ \ L_1,  \ \ L_3,  \ \ L_4,  \ \ \alpha, \ \  \beta, \ \ r_1 \ \ and \ \ \chi$ are defined by
  \ba \Omega \A  \stackrel
{\rm def.}{=} \A r^2+{L_1}^2\; , \qquad  \Upsilon \stackrel {\rm
def.}{=} r^2+a^2-2Mr-L^2
  \; , \quad L_1 \stackrel
{\rm def.}{=} L+a\cos\theta,  \nonu
 L_3 \A \stackrel {\rm def.}{=} \A aL\cos^3\theta(L^2+r^2+a^2)+\cos^2\theta(L^4+L^2Mr-a^2rM+L^2r^2-L^2a^2+Mr^3)\nonu
 \A \A+aL\cos\theta(a^2+r^2-4rM-3L^2)+a^2L^2+a^2Mr
 -L^4-r^3M-rML^2-r^2L^2,\nonu
  L_4 \A \stackrel {\rm def.}{=} \A La^2\cos\theta(\cos^2\theta-3)-a^3\cos^2\theta+arM\cos^2\theta+3aL^2\cos^2\theta-rM\chi-r^2a-2aL^2,\nonu
  L_5 \A \stackrel {\rm def.}{=} \A -a{\cal B }\cos^3\theta(a\cos\theta+4L)-\cos^2\theta\Biggl(r^4+8a^2L^2+6r^2L^2-a^4-3L^4-8MrL^2+4Mra^2\Biggr)\nonu
  \A \A +4aL{\cal B }\cos\theta +2Mra^2+L^4+r^4+r^2a^2+2r^2L^2+3a^2L^2, \quad
\alpha  \stackrel {\rm def.}{=} r_1\cos \phi+a \sin\phi, \nonu
\chi \A \stackrel {\rm def.}{=} \A  a \sin^2 \theta-2L\cos\theta,
\qquad  \beta \stackrel {\rm def.}{=} r_1 \sin \phi-a \cos \phi\;,
\quad r_1 \stackrel {\rm def.}{=}
\sqrt{r^2+L(L+2a\cos\theta)}.\nonu
 \A \A\ea
Using \be \left( {h^\alpha}_i \right)
=\left({\Lambda^\alpha}_\gamma\right)
 \left( {h^\gamma}_i \right)_1, \ee  in Eq. (10) to calculate the non-vanishing components needed to the calculations of energy,  we finally
 get\footnote{ Science the  equations of
 motion (7) is just Einstein equations  written in terms of tetrad fields $h^\alpha_i$ therefore, Eq. (34) is an exact solution to Eq.
 (7). This case is studied intensively by  Hayashi and Shirafuji (cf., Ref. 12) Eqs. (7$\cdot$ 2)$\sim$(7$\cdot$ 11) and references therein).}
 \ba
\Sigma^{1 0 1}\A \cong
\A\frac{a}{2r^5}\left([aL\cos\theta+Mr+L^2]a\sin^2\theta-4aL^2\cos^2\theta+[L^2+Mr]2L\cos\theta\right)+O\left(\frac{1}{r^6}\right),\nonu
\Sigma^{2 0 1}\A \cong \A
\frac{-a}{2r^5\sin\theta}\left([Ma\cos\theta-Lr]\sin^2\theta-2LM\cos^2\theta\right)+O\left(\frac{1}{r^6}\right),\nonu
\Sigma^{3 0 1}\A \cong \A
\frac{-L}{2r^5\sin^2\theta}\left(a^2\cos\theta\sin^2\theta+aL\sin^2\theta-L^2\cos\theta\right)+O\left(\frac{1}{r^6}\right),\nonu
\Sigma^{4 0 1}\A \cong
\A-\frac{1}{2r^4}\left(2a^2r\cos^2\theta+a^2M\sin^2\theta+2aL\cos\theta(2r-M)+2r[L^2-r^2]\right)+O\left(\frac{1}{r^5}\right).\nonu
\A \A  \ea

Using Eq. (35) in Eq. (9) we finally obtain\footnote{We introduce
${{\Sigma}^{\mu 0 k}}_{M=0,a=0,L=0}$, in Eq. (36), to remove the
divergence appearers from term like $r$. It is worth to mention
that we cannot  use the expression ${{\Sigma}^{\mu 0 k}}_{r
\rightarrow \infty}$ because the spacetime we use is not
asymptotically flat.} \ba \A \A P^{(0)}=E=-\oint_{S \rightarrow
\infty}
 dS_k \Pi^{(0) k}=-\displaystyle {1  \over 4 \pi} \oint_{S \rightarrow \infty}
 dS_k  \sqrt{-g}
{h^{(0)}}_\mu \left({\Sigma}^{\mu 0 k}-{{\Sigma}^{\mu 0
k}}_{M=0,a=0,L=0} \right)\nonu
\A \A \cong
M+\frac{L^2}{r}-\frac{L^2M}{r^2}-\frac{L^2(5a^2+6L^2)}{6r^3}+O\left(\frac{1}{r^4}\right).\ea
Eq. (36) is a satisfactory results.$^{ \rm {33)}}$ It is clear
from (36) that the energy content is shared by both  the interior
and exterior of the Kerr-NUT spacetime. The total energy when $r
\rightarrow \infty$ gives the ADM (Arnowitt-Deser-Misner).

The  non vanishing components needed to calculate the spatial
momentum have
 the form\footnote{Terms like $M^2$, $L^3$, $L^3M$,$ M^2a$, $\cdots$ ect. are neglected in this calculations.}
\ba   \Sigma^{(1) 0 1}\A \cong \A\frac{-\sin^2\theta\cos\phi}{8\pi
r}\left(2Mr+2L^2+3aL\cos\theta\right)+O\left(\frac{1}{r^2}\right),\nonu
 \Sigma^{(2) 0 1}\A \cong \A \frac{-\sin^2\theta\sin\phi}{8\pi r}\left(2Mr+2L^2+3aL\cos\theta\right)+O\left(\frac{1}{r^2}\right),\nonu
 \Sigma^{(3) 0 1}\A \cong \A \frac{-\sin\theta}{8\pi r}\left(3aL\cos^2\theta+2L^2\cos\theta+2Mr\cos\theta-aL\right)+O\left(\frac{1}{r^2}\right).\ea
Using Eq. (37) in Eq. (9), we finally get
 the spatial momentum in the
form \be P_1=-\oint_{S \rightarrow \infty}
 dS_k \Pi^{(1) k}=-\displaystyle {1  \over 4 \pi} \oint_{S \rightarrow \infty}
 dS_k  e {\Sigma}^{(1) 0 k} =0 ,\quad by \ same \  method
\ P_2=0, \quad P_3\cong\left(\frac{1}{r^2}\right).\ee

Now turn our attention to the calculation of angular-momentum. The
non vanishing components of the angular-momentum are given by \ba
M^{(0)(1)}(r,\theta,\phi) \A \A \cong
 \frac{-M\cos^2\theta}{4\pi r^2}(\{L^2+rM+aL\cos\theta+2M^2\}\cos\phi+aM\sin\phi)+O\left(\frac{1}{r^2}\right), \nonu
  M^{(0)(2)}(r,\theta,\phi) \A \A  \cong
 \frac{-M^2\sin\phi\cos^2\theta}{4\pi r}+O\left(\frac{1}{r^2}\right), \qquad
M^{(0)(3)}(r,\theta,\phi) \cong
\frac{M^2\sin\theta\cos\theta}{4\pi
r}+O\left(\frac{1}{r^2}\right),\nonu
M^{(1)(2)}(r,\theta,\phi)\A \A \cong
\frac{-M^2a\sin\theta\cos^2\theta}{4\pi
r^2}+O\left(\frac{1}{r^3}\right),\nonu
M^{(1)(3)}(r,\theta,\phi) \A \A   \cong
\frac{M\cos\theta([r+M]\cos\phi+a\sin\phi)}{4\pi
r}+O\left(\frac{1}{r^2}\right),\nonu
 M^{(2)(3)}(r,\theta,\phi) \A \A \cong \frac{M\cos\theta([r+M]\sin\phi-a\cos\phi)}{4\pi r}+O\left(\frac{1}{r^2}\right).\ea
Using Eq. (39) in (11) we get
 \be L^{(0)(1)} = {\int_0^\pi}{\int_0^{2\pi}}{\int_{0}^\infty}
 d\theta d\phi dr \left[M^{(0)(1)} \right] =
0,\ee which is a consistent  results.
  By  the same  method  we  finally  obtain \be
 L^{(0)(2)} =  L^{(0)(3)} =
 L^{(1)(2)}=L^{(1)(3)} = L^{(2)(3)}= 0.\ee

We show by explicit calculations that the energy-momentum tensor
which is a coordinate independent does not give  a consistent
result of the angular momentum when applied to the tetrad field
given by Eq. (12)!

To show if Eq. (34) continue to give acceptable result of energy
we use the superpotential (26). The coframe of Eq. (34) takes the
form  \be
{{{\vartheta^{{}^{{}^{\!\!\!\!\scriptstyle{'}}}}}{_{}{_{}{_{}}}}}^{\alpha}}
=\left({\Lambda^\alpha}_\gamma\right) {\vartheta_1}^{\gamma}, \ee
with $\left({\Lambda^\alpha}_\gamma\right)$ and
${\vartheta_1}^{\gamma}$ are given by Eqs (28) and (31)
respectively.
  If we take coframe  (42), as
well as the trivial Weitzenb$\ddot{o}$ck connection, i.e.,
${\Gamma^\alpha}_\beta=0$  and substitute into (26) we finally get
the temporal component of the translation momentum in the form \ba
\widetilde{H}_{\hat{0}} \A \A \cong -\frac{\sin \theta}{8r^2\pi}
\left[3a^2M\cos^2\theta+6aLM
\cos\theta-2arL\cos\theta+2ML^2-2rL^2-a^2M+2r^3-2r^2M
\right]d\theta\wedge d\phi\nonu
\A \A +\cdots \cdots+O\left(\frac{1}{r^3}\right).\ea Computing the
total energy up to order $O\left(\frac{1}{r^2}\right)$ at a fixed
time in the 3-space with a spatial  2-dimensional boundary surface
$\partial S =\{r = R, \theta,\phi\}$ we obtain\footnote{We
introduce $\Biggl\{\widetilde{H}_{\hat{0}}\Biggr\}_{M=0,a=0,L=0}$
to remove the divergence appearers from term like $r$. It is worth
to mention that we cannot  use the expression
$\Biggl\{\widetilde{H}_{\hat{0}}\Biggr\}_{r \rightarrow \infty}$
because the spacetime we use is not asymptotically flat.} \be
\widetilde{E} =\int_{\partial S}
\left(\widetilde{H}_{\hat{0}}-\Biggl\{\widetilde{H}_{\hat{0}}\Biggr\}_{M=0,a=0,L=0}\right)
\cong M+\frac{L^2}{R}-\frac{L^2M}{R^2}
+O\left(\frac{1}{R^3}\right),\ee which is the energy of Kerr black
hole when $L=0$$^{ \rm {33)}}$.

 The  necessary components needed to calculate the spatial momentum
 $\widetilde{H}_{\hat{\alpha}}, \
\hat{\alpha}=1,2,3$ have the following  components  \ba
\widetilde{H}_{\hat{1}} \A=\A \frac{2\cos\phi
\sin^2\theta[3aL\cos\theta+2L^2+2Mr+4M^2]}{r}d\theta\wedge
d\phi+\cdots \cdots, \nonu
\widetilde{H}_{\hat{2}} \A=\A \frac{2\sin\phi
\sin^2\theta[3aL\cos\theta+2L^2+2Mr+4M^2]}{r}d\theta\wedge
d\phi+\cdots \cdots, \nonu
\widetilde{H}_{\hat{3}}\A=\A  \frac{2\sin
\theta\left([3aL\cos\theta+2L^2+2Mr+4M^2]\cos\theta-aL\right)}{r}d\theta\wedge
d\phi+\cdots \cdots. \ea Using Eqs. (45) in (26), we finally get
the spatial momentum in the form \be P_1=P_2=P_3\cong
O\left(\frac{1}{R^2}\right).\ee
\newsection{Second Kerr-NUT spacetime}
The covariant form of the second Kerr-NUT tetrad field having
axial symmetry in spherical coordinates, can be written as
 \ba
\left( {h^\alpha}_i \right)_2\A=\A \left( \matrix{ {\cal A}_1 & 0
& 0 & 0\vspace{3mm} \cr -\sin\phi {\cal A}_4  & \sin\theta\cos\phi
{\cal A}_2&  \cos\theta\cos\phi{\cal A}_3 &- \sin\theta\sin\phi
{\cal A}_5 \vspace{3mm} \cr \cos\phi {\cal A}_4  &
\sin\theta\sin\phi {\cal A}_2&  \cos\theta\sin\phi{\cal A}_3
&\sin\theta\cos\phi {\cal A}_5  \vspace{3mm} \cr  0 & \cos\theta
{\cal A}_2&  -\sin\theta {\cal A}_3  & 0 \cr } \right)\nonu
\A \A \ \ where  \ \ {\cal A}_i, \  \ i=1 \cdots 5 \ \ are \ \
defined  \ \ by \ \  Eq. \ \ (13).  \ea
  Tetrad field (47) has the same associated metric of  tetrad (12), i.e.,
   Kerr-NUT spacetime given by Eq. (14). Tetrad (47) is related to tetrad (12)
   through the relation \be \left( {h^\alpha}_i \right)_2=\left({{\Lambda_1}^\alpha}_\gamma\right) \left( {h^\gamma}_i
   \right)_1,\ee
  where $\left({{\Lambda_1}^\alpha}_\gamma\right)$ is the local Lorentz transformation given
  by\footnote{Eq. (48) is an exact solution to the equations of
  motion (7) due to the reasons discussed for Eq. (34).}
   \be
\left({{\Lambda_1}^\alpha}_\gamma\right)  \stackrel {\rm def.}{=}
\left( \matrix{ 1 &  0 & 0 & 0 \vspace{3mm} \cr  0  &  \sin\theta
\cos\phi &  \cos\theta \cos\phi & - \sin\phi \vspace{3mm} \cr 0  &
\sin \theta \sin \phi& \cos\theta \sin\phi & \cos\phi \vspace{3mm}
\cr 0  & \cos\theta & -\sin\theta  & 0 \cr }\right)\; ,\ee
Following the same technics used in \S 3 to calculate energy we
finally get the same non-vanishing components of $\Sigma^{a 0 1}$
asymptotically as given by Eq. (15) and the form of energy will be
the same as given by Eq. (16).  Repeat the same calculation done
in \S 3, in which  Riemannian connection  1-form is employed and
using tetrad (47), we obtain same divergence value of energy given
by Eq. (30).

To remove such unphysics results, let us consider the local
Lorentz transformation described by Eq. (31) and follow the same
procedure done in the previous section we can remove the
divergence appears.

\newsection{ Discussion and conclusion }

 The tetrad field given by Eq. (12) whose  associated metric gave the Kerr-NUT spacetime is used. From this spacetime one can
  generate Kerr and Schwarzschild  spacetimes by putting $L=0$ and $a=0, \ L=0$ respectively.  Our results can be summarized as follows:\vspace{0.4cm}\\
   $\bullet$ The energy content of tetrad field (12) using  the gravitational energy-momentum tensor (9) was calculated
     and  a divergent result was obtained.
    It is worth  to mention that the value of energy did not depend on the radial coordinate $r$, i.e.,  the divergence
 is not related to the fact that when $r\rightarrow \infty$ the value of energy will be infinity. The source of divergence is related to the
 NUT parameter $L$, because when $L$  is nil the value of energy becomes finite and  will
 be identical with the ADM which is the energy of Kerr [Ref. 46)  Eq. (25)].\vspace{0.4cm}\\
   $\bullet$ To assure that the divergence of  energy is not related to the expression which has been
    used for calculation, another expression to recalculate the energy has been applied, i.e.,
    we have used the Riemannian connection 1-form, and have got the same divergent
     result.\vspace{0.4cm}\\
   $\bullet$ To remove such an unacceptable result, the local Lorentz transformation  (31) has been  suggested.
    This transformation when is multiplied by tetrad (12)  and by repeating the same calculation of energy, using the gravitational energy-momentum tensor,
     we  got the correct value of Kerr-NUT spacetime [Ref.  47) Eq. (44)]. Tetrad (34) gave a very satisfactory result of the spatial
     momentum, but for the angular momentum the result is not  correct. We may claim  that the unfamiliar result of the angular momentum are related to
     the expression  used in calculation, Eq. (11).  To assure that the result of energy is correct,  the Riemannian
     connection 1-form (26) has been used and  the same consistent result have been reached,  Eq. (44).$^{ \rm {33)}}$ \vspace{0.4cm}\\
   $\bullet$ We have used another tetrad given by Eq. (47) and have calculated the energy using the two expressions: the gravitational energy-momentum
   Eq. (9) and the Riemannian connection 1-form (26) and got the same result, i.e., divergent value, which is given by Eq. (16).
   The divergence of the energy is due to the fact
   that tetrad (47) is related to tetrad (12) through the local Lorentz transformation (48). It is of interest to note that
   in the spherical symmetric
   case, i.e., when $L=0$ and $a=0$ is shown in$^{ \rm {27)}}$ that local Lorentz transformation (48) when applied to Eq. (12)
   the expression of energy
   became finite [Ref.  27) \ \ Eq. (4$\cdot$ 13)]. However, this procedure did not work here because  an axially symmetric tetrad is used. \vspace{0.4cm}\\
   $\bullet$ When local Lorentz transformation (31) is multiplied  by Eq. (47) combined with a repetition of the same procedure   to
    calculate the energy,  a finite and consistent result have been obtained.\vspace{0.4cm}\\
   $\bullet$ It is of interest to note that expression (11) did not give   the correct result of angular momentum.
  In our forthcoming research work, calculations of the total conserved quantities related to tetrad fields (12)
        and (47) will be dealt with.

\bigskip
\bigskip
\centerline{\Large{\bf Acknowledgements}}

The author would like to thank Professor A. Hussain; English
Department college of Arts, King Faisal University, Saudi Arabia
for taking the time in reviewing the langauge of the manuscript.

\end{document}